\documentclass[journal, onecolumn]{IEEEtran}
\usepackage[english
]{babel}
\usepackage[utf8]{inputenc}
\usepackage[T1]{fontenc}
\usepackage{epsfig}
\addto\extrasenglish{\languageshorthands{english}}
\renewcommand{\baselinestretch}{2.0}

\usepackage{graphicx}
\usepackage{float}
\usepackage{epstopdf}

\usepackage{color}
\usepackage{xcolor}
\usepackage{colortbl}

\usepackage{amssymb, amsmath}
\usepackage{mathtools}

\usepackage{algorithm2e}
\usepackage{algpseudocode}

\usepackage{booktabs}

\usepackage{listings}

\usepackage{courier}
\usepackage{tikz}
\usepackage{pgfplots}

\begin{document}
\title{Performance Limits of a Cloud Radio}

\author{
Maaz~M.~Mohiuddin\authorrefmark{1},~Varun~Maheshwari\authorrefmark{1},~Sreejith~T.~V.\authorrefmark{1},\\~Kiran~Kuchi\authorrefmark{1},~G.~V.~V.~Sharma\authorrefmark{1}~and~Shahriar~Emami\authorrefmark{2}

\thanks{\authorrefmark{1} Authors (\{ee09b015, ee09b035, ee12p002, kkuchi, gadepall\}@iith.ac.in) are with the Department of Electrical Engineering, Indian Institute of Technology Hyderabad,
Hyderabad, India. This work was funded by Samsung, South Korea.
The authors would like to thank Dr. Radhakrishna Ganti, Department of Electrical Engineering at Indian Institute of Technology Madras for his inputs on Point Processes.
}

\thanks{\authorrefmark{2} Dr. Shahriar Emami (shahriar.e@sisa.samsung.com) is with Samsung R \& D, 75 Plumeria Drive, San Jose, CA, 95134}
}
\maketitle

\begin{abstract}
Cooperation in a cellular network is seen as a key technique in managing other cell interference to observe a gain in achievable rate. In this paper, we present the achievable rate regions for a cloud radio network using a sub-optimal zero forcing equalizer with dirty paper precoding. We show that when complete channel state information is available at the cloud, rates close to those achievable with total interference cancellation can be achieved. With mean capacity gains, of up to $200~\%$ over the conventional cellular network in both uplink and downlink, this precoding scheme shows great promise for implementation in a cloud radio network. To simplify the analysis, we use a stochastic geometric framework based of Poisson point processes instead of the traditional grid based cellular network model.

We also study the impact of limiting the channel state information and geographical clustering to limit the cloud size on the achievable rate. We have observed that using this zero forcing-dirty paper coding technique, the adverse effect of inter-cluster interference can be minimized thereby transforming an interference limited network into a noise limited network as experienced by an average user in the network for low operating signal-to-noise-ratios. However, for higher signal-to-noise-ratios, both the average achievable rate and cell-edge achievable rate saturate as observed in literature. As the implementation of dirty paper coding is practically not feasible, we present a practical design of a cloud radio network using cloud a minimum mean square equalizer for processing the uplink streams and use Tomlinson-Harashima precoder as a sub-optimal substitute for a dirty paper precoder in downlink.
\end{abstract}


\IEEEpeerreviewmaketitle

\section{Introduction}
The evolution of cellular communication networks from 1G through 4G \cite{3p9G:standard}
has resulted in a steady increase in the allowable rates for users (UEs) in the interior of the cell boundary. However, due to other cell interference (OCI) the UEs at the edge of the cell experience reduced rates. Mitigation of OCI using interference cancellation strategies results in improvement of allowable rates for all UEs, cell-edge UEs in particular. This can be achieved by a cloud radio network, which is also labeled as network multiple-input-multiple-output (MIMO) \cite{Foschini:Mar98}\nocite{Telatar:June95}\nocite{Foschini:2006}\nocite{Poor:2009}\nocite{LTEcomp}\nocite{Gorokhov}\nocite{Fettweis}\nocite{Gaal:2012}\nocite{Foschini:2006a}\nocite{Gersbert:Dec2010}\nocite{Wang:Nov2011}\nocite{Poor:2012a}\nocite{Lozano:2009e}\nocite{Hottinen:June2009}\nocite{Peel:Jan05}\nocite{Peel:Mar05}-\cite{Caire:2009A} centralized precoding architecture in literature. In a cloud radio network, a group of near by base-stations (BSs) function in cooperation to cancel the interference due to each other. In the downlink, the cloud, which is a collection of multiple BSs, serves a group of UEs simultaneously with multiple streams. These streams are appropriately precoded, so that interference-free decoding is possible at the receiver end. In the uplink, the cloud receives multiple streams from each UE, a stream corresponding to each point to point channel between a BS and a UE. These streams are jointly decoded at the cloud to nullify interference effects.

The improvements in rate in a cloud radio come at the expense of an increase in BS and UE complexity. Hence, it is important to quantify the capacity of a cloud radio to understand the feasibility of employing such an architecture. The channel capacity under a certain encoding-decoding strategy i.e. the achievable rate is characterized by computing the post processing signal-to-interference-plus-noise-ratio (SINR). Traditionally, the SINR metric for cellular networks is obtained using the hexagonal grid based model, where the BSs are placed in the form of a grid and the UEs are scattered in the area. In \cite{heath}, the authors have used the grid based model to lay the fundamental limits on the capacity of a cloud radio network. The authors have suggested that even with a faster back haul or more efficient signal processing, the gain in capacity from a cloud radio network as opposed to a conventional cellular network cannot be improved due to effects of inter-cluster interference \cite{Mennerich:2011}\nocite{Gesbert:2008icc}-\cite{Garcia:2010}. However, the high theoretical gain in achievable rate through a cloud radio network calls for an exploration for the region of practically achievable gain. Furthermore, this analysis technique suffers from serious tractability issues and very high simulation time. The increased complexity of a cloud radio network as opposed to a conventional cellular network renders the grid based model further unfit for analysis. In this paper, we use a tractable approach to evaluate the rate as discussed in \cite{andrews}. We modify the Poisson point process (PPP) based stochastic geometric framework \cite{Baccelli:1997}\nocite{Haenggi:2009}-\cite{Stoyan} designed for a conventional cellular network to fit a cloud radio network.

Besides characterizing the capacity region of a cloud radio, it is also vital to explore practically realizable transmitter and receiver configurations. In this paper, we discuss the evaluation of the capacity region \cite{tse}\nocite{tse}\nocite{Wu:2004}-\cite{Sriram:2003} of a cloud radio under sub-optimal encoding-decoding strategy and present a practically realizable system design which achieves this capacity. The major contributions of this paper can be summarized as follows:
\begin{enumerate}
\item Characterization of the capacity region of a cloud in downlink with equal power constraints
\item Characterization of the capacity region of a cloud in uplink equal power constraints
\item Evaluation of theoretical bounds on the capacity region
\item Characterization of the capacity region of a cloud under partial channel state information (CSI) and geographic clustering
\item Practically feasible cloud radio design using Tomlinson Harashima precoder (THP).
\end{enumerate}

The rest of the paper is organized as follows. In Section II, we present the downlink system model and discuss the duality between downlink and uplink channel capacity regions in a cloud radio network. Section III, details the evaluation of the downlink channel capacity using a sub-optimal zero forcing- dirty paper coding (ZF-DPC) based receiver \cite{shitz}. We also evaluate the theoretical bounds on the capacity under total interference cancellation and spatial match filter precoding. In Section IV, we evaluate the uplink capacity of a cloud radio under ZF-DPC decoding. We also evaluate the capacity of a cloud radio under a low-complexity  minimum mean square error (MMSE) receiver \cite{Cioffi:Stanford}, \cite{Winters:84}.

In Section V, we discuss the effect of limited channel feedback on the capacity region. We also discuss the evaluation of theoretical bounds on capacity of a cloud radio with limited channel feedback. Then, we cluster geographically close BSs to form a cloud and evaluate the capacity of the network under limited channel feedback from only intra-cluster BSs. Geographical clustering with limited feedback is the closest approximation to a real life cloud radio network where the cluster size is limited by the relative distances between the BSs and the amount of feedback is restricted due to infra-structural limitations. In \cite{heath}, the authors have shown that the achievable rate of such a cloud radio network scales with the signal-to-noise-power-ratio (SNR) at low operating SNR and the achievable rate saturates at high SNR. We have observed that a ZF-DPC based cloud radio network shows similar performance, thereby making it noise limited at low, medium SNRs and interference limited at high SNRs. Section VI details a practically realizable system design using a Tomlinson Harashima precoder \cite{thp}, a simpler, sub-optimal version of dirty paper decoder. We conclude with future scope in Section VII.

\section{System Model}
In this section, we present the system model for a cloud radio operating in downlink. Then we relate the downlink system model to the system model for a cloud radio operating in uplink using the duality between the uplink and downlink in a cellular network.

\subsection{Downlink System Model}
We follow the system model as in \cite{andrews} with modifications to incorporate interactions between multiple BSs and multiple UEs. The BSs follow a uniform PPP $\Phi_b$ with an intensity $\lambda_b$ and the UEs follow an independent uniform PPP $\Phi_u$ with an intensity $\lambda_u$ in Euclidean space. The cloud radio network is modelled as a collection of all the BSs in the PPP space. The BSs are indexed as BS$_i$, where $i~\in~\left\lbrace 1,2,\cdots, | \Phi_b |\right\rbrace$ and the UEs are indexed as UE$_j$, where $j~\in~\left\lbrace 1,2,\cdots, | \Phi_u |\right\rbrace$. Each UE is associated with a primary BS which is the geographically closest BS to itself and all other BSs in the cloud are the secondary BSs to that UE. The UEs are served in a round robin fashion by grouping $k$ UEs, $k$ being the size of the cloud, such that each UE has a different primary BS. To compare the downlink channel capacity of a cloud radio network with that of a set of downlink cellular network links between $k$ BSs and $k$ UEs, it is important to ensure that each BS has an associated UE in every round robin cycle. This is ensured by taking $\lambda_u$ to be sufficiently larger than $\lambda_b$, so that each BS has atleast one UE for primary association.

The data stream from BS$_i$ to UE$_j$ is transmitted with a constant transmit power of unity and experiences Rayleigh fading with mean power $1/\mu$. A standard path loss propagation model is used with path loss exponent $\alpha = 4$, i.e. path loss propagation model for urban areas. Therefore, the received signal voltage for a stream from BS$_i$ to UE$_j$ is given by $h_{ij}z_{ij}^{-\alpha/2}$, where $z_{ij}$ is the distance between BS$_i$, UE$_j$ and $h_{ij}$ is the corresponding Rayleigh fade. The noise across all streams is assumed to be additive white Gaussian noise (AWGN). In the PPP framework, the notion of signal-to-noise-ratio (SNR) is in the average sense for the entire network. For a noise power of $\sigma_i^2$ and a BS intensity of $\lambda_b$, the average SNR is $1/(16 \lambda_b^2 \sigma^2_i)$ \cite{andrews}.

A conventional cellular network is a collection of several point to point links. In a cloud radio network, several BSs and UEs cooperate to reduce interference and increase the achievable rate. The total number of BSs is referred to as the size of the cloud ($k$) and is numerically equal to the number of UEs a cloud serves at any instant. Hence, the received signal vector in downlink is given as:

\begin{equation}
{Y_{k\times 1}} = {H_{k\times k}}~{X_{k\times 1}} + {N_{k\times 1}} \label{sys}
\end{equation}
where, ${Y = [y_1~y_2~\hdots~y_k]^\dagger}$ is the received signal vector, \[
{H} = \left[ \begin{array}{cccc}
h_{11}z_{11}^{-\alpha/2} & h_{12}z_{12}^{-\alpha/2} & \hdots & h_{1k}z_{1k}^{-\alpha/2}\\
h_{21}z_{21}^{-\alpha/2} & h_{22}z_{22}^{-\alpha/2} & \hdots & h_{2k}z_{2k}^{-\alpha/2}\\
  \vdots & \vdots & \vdots & \vdots\\
h_{k1}z_{k1}^{-\alpha/2} & h_{k2}z_{k2}^{-\alpha/2} & \hdots & h_{kk}z_{kk}^{-\alpha/2}
 \end{array} \right] \],\\${X = [x_1~x_2~\hdots~x_k]^\dagger}$ is the transmitted symbol vector and
${N = [n_1~n_2~\hdots~n_k]^\dagger,~n_i}\sim \mathcal{CN}{(0,\sigma_i^2)}$ is AWGN.

Upon closer analysis, it can be seen that the channel matrix ${H}$ of a cloud radio network in downlink represents a vector Gaussian broadcast channel (BC). Next, we discuss the established duality between the capacity regions of the vector Gaussian multiple access channel (MAC) and a vector Gaussian BC.

\subsection{Uplink-Downlink Duality}
A cloud radio network resembles a vector Gaussian BC in downlink and a vector Gaussian MAC in the uplink. According to the Uplink-Downlink duality principle \cite{tse}\nocite{tse}\nocite{Wu:2004}-\cite{Sriram:2003}, the achievable sum rate for a network with flat fading and AWGN channels characterized by the matrix ${H}$, is same for both uplink and downlink; and is given by:
\begin{equation}
C_{sum} = \sup_{{D} \in \mathbb{A}}~{\log}~|{I} + {H} {D} {H^\dagger}| \label{csum}
\end{equation}
where $\mathbb{A}$ is the set of matrices $\mathbf{D}$ with $Tr[\mathbf{D}] \leq P$, $\mathbf{D}$ being the signal covariance matrix ${E[xx^\dagger]}$. Furthermore, by the point-to-point reciprocity developed in \cite{tse}, for a system ${Y}={HX+N}$ and its dual system ${Y}={H^T X+N}$, the set of achievable SINR's are same when the sum total of the transmit powers is same in uplink and downlink. In this paper, we use the aforementioned results to evaluate the capacity region for a cloud radio in downlink under the constraint of equal power for all transmitters and use the same capacity region for the cloud radio operating in uplink.

\section{Downlink Channel Capacity}
When CSI is known at the transmitter, spatial multiplexing techniques can be used to obtain high rate. In the downlink, the cloud transmits multiple streams to multiple UEs. However, as the receivers cannot cooperate, the achievable rate depends on the cloud's capacity to orthogonally precode the streams. In \cite{dpc}, Costa proved that by dirty paper (DP) precoding, in a network where the interference is non-causally known at the transmitter, it is possible to achieve the
same capacity as if there were no interference. However, the complexity of an optimal DP precoder can be problematic for online implementation. In \cite{shitz}, the authors have presented a reduced-complexity precoder with sub-optimal performance. This technique uses QR factorization of the channel matrix to obtain a lower triangular interference matrix which can be decoded using DPC with ease. This technique nulls the interference between data streams and hence the name zero forcing dirty paper coding. We have used the reduced complexity sub-optimal ZF-DPC based algorithm in to find the achievable rate region of a cloud radio network.

\subsection{ZF-DPC}
This is the main result of the paper. The proof of the encoding and decoding strategy is discussed in Appendix A. The key points in the proof are:
\begin{enumerate}
\item The channel matrix ${H}$ is such that a diagonal element is greater than all other elements in the corresponding row and corresponding column. This structure in ${H}$ is due to the association rule of a UE with its geographically closest BS as discussed in the system model. Such a $H$ matrix ensures that the elements of the $R$ matrix after factorization are such that the diagonal elements are the greatest.

\item The precoding matrix W is chosen as ${Q^\dagger}$ because ${Q}$ being unitary (${QQ^\dagger = I}$), the received signal symbols are obtained in an upper triangular form which can be handled by a reduced complexity DP precoder.

\item A full complexity DP precoder achieves interference cancellation by performing precoding operation on a full channel matrix. In our approach, complexity reduction is achieved in DPC by eliminating half the interference terms through the simple decomposition in Step 2.

\item For the sub-optimal ZF-DPC being used, the transmit symbol power is chosen to be 1. Therefore the achievable rate for UE$_i$ is given by \eqref{cpdc}. Based on our association rule, the matrix $R$ has the greatest elements on the diagonal thereby maximizing the capacity of the ZF-DPC based cloud radio network.
\begin{equation}
{C_i} = {\log\left( 1+ \frac{|r_{ii}|^2|x_i|^2}{\sigma_i^2} \right)} bps/Hz \label{cpdc}
\end{equation}
\end{enumerate}

\begin{figure} [thb]
    \centerline{
        \epsfig{figure=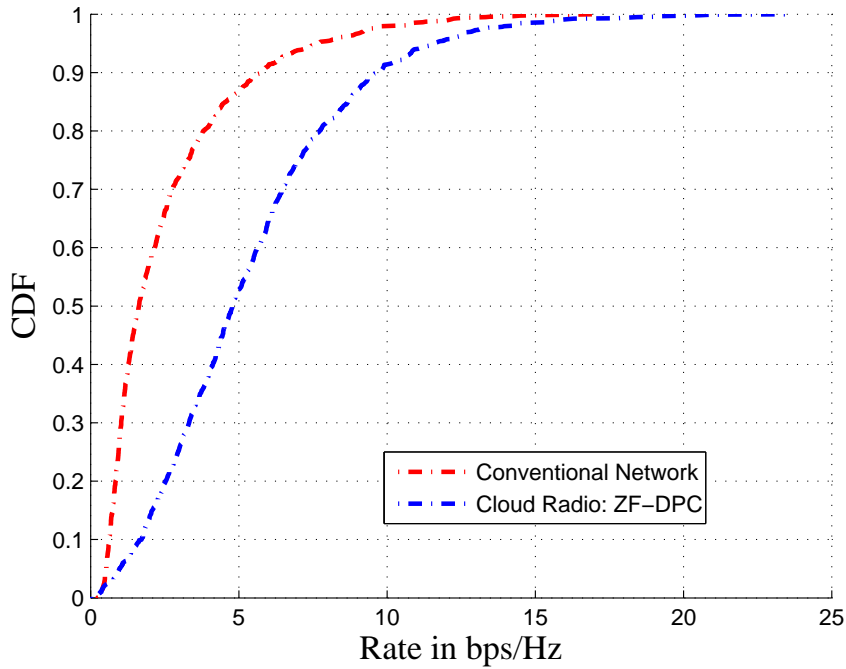, width=5in}
    }
        \vspace*{0.0cm}  \caption{Rate statistics of a conventional cellular network and a cloud radio network with ZF-DPC; operating $SNR~= 10~dB$} \label{conv_zf}
\end{figure}

We have simulated a cloud radio network with ZF-DPC encoding using a PPP framework to obtain the statistics of the achievable rate. We use a rectangular area of $10 Km$ with $\lambda_b = 0.3$, which gives the average number of BSs in the cloud to 30. We have chosen a cloud area network with 30 BSs so as to balance the trade-off between simulation complexity and adverse effects of a small cloud on the capacity statistics. The individual SNR at the transmitter for each BS in the cloud $(1/ \sigma^2_i)$ is taken to be $10~dB$. Fig \ref{conv_zf} shows the comparison between the CDF of rate for a conventional cellular network \cite{andrews} and a cloud radio network. It is observed that the mean rate for a network has increased by $202~\%$ from $1.63~bps/Hz$ to $4.93~bps/Hz$. The cell edge UEs in a network are characterized by the $0.05~\%$ point on the rate CDF curve. The cell-edge UEs enjoy almost double the rate as observed in a conventional cellular network i.e. from $0.51~bps/Hz$ for a conventional cellular network to an achievable rate of $0.97~bps/Hz$ for a cloud radio network operation.

\begin{figure}[!t]
\centering
\includegraphics[width =5in]{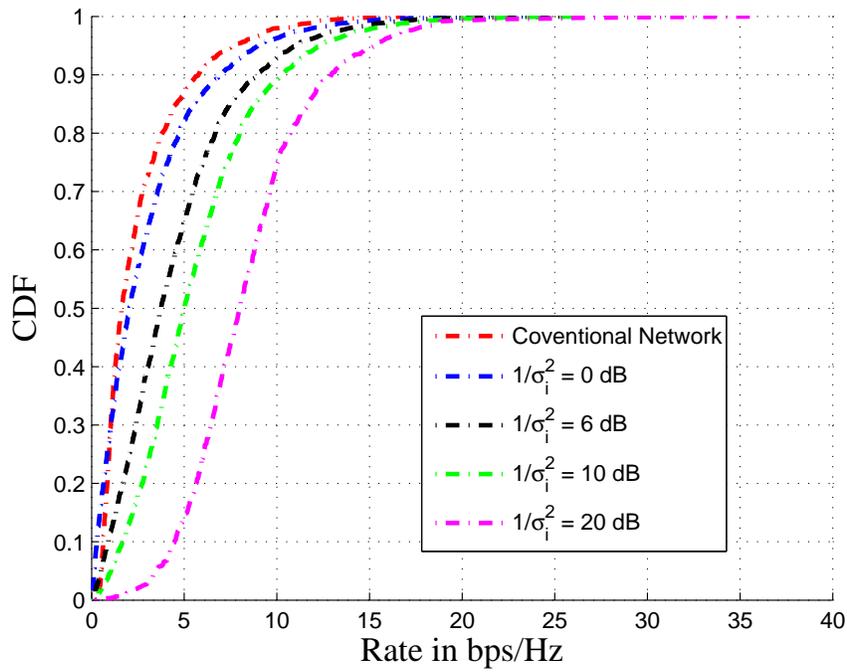}
\caption{Variations in rate statistics as a function of noise level}
\label{noise}
\end{figure}

As a ZF-DPC encoder achieves interference cancellation, the cloud radio network under analysis is no longer interference limited like a conventional cellular network i.e. the capacity of the network scales with the decrease in noise power. Fig \ref{noise} shows the variations of the rate statistics with a change in the noise power. Clearly, the shift in the CDF curve to the right with the decrease in noise power shows that the network is noise limited as opposed to interference limited. Therefore, by boosting up the transmit power at the BSs, an increase in capacity can be realized. The mean achievable rates for a cloud radio networks operating at BS SNR of $0~dB$, $-6~dB$, $10~dB$ and $20~dB$ are $2.1,~3.746,~4.93,~7.792~bps/Hz$ respectively.  For $1/ \sigma_i^2~=20~dB$, the cell-edge UE observes an achievable rate of $3.46~bps/Hz$ which is $270~\%$ higher than that at $10~dB$ and almost 7 times of the cell-edge rate achievable through a conventional cellular network.

A further improvement in capacity can be achieved by power-loading using iterative water-filling. However, the gain from a constrained ZF-DPC based cloud radio network is already so high, that the minor improvements in rate through water-filling at the cost of increasing the complexity of the system have a degrading effect on the entire system design. Therefore, we choose not to explore the slight increase in rate achievable through water filling.

To evaluate the performance of the ZF-DPC based cloud radio, we have developed theoretical bounds on the capacity of a cloud radio with total interference cancellation and interference suppression through match filter combining of multiple streams as discussed in the next section.

\subsection{Theoretical Bounds on Capacity}
A ZF-DPC sub-optimally processes the OCI streams to obtain interference cancellation and convert the OCI steams in useful signal. In a conventional network, if the OCI were to be cancelled without converting it to useful signal we obtain a lower bound on the ZF-DPC capacity. On the other hand, if the interference streams are combined using a match filter to add to the signal power, then an upper bound on the capacity of ZF-DPC based cloud radio is obtained. The details of these bounds are discussed in the following segments.

\subsubsection{Total Interference Cancellation Bound}
In the conventional cellular network, OCI reduces the rate observed by cell-edge UEs. To increase the rate observed by cell-edge UEs, OCI needs to be suppressed. Consider a downlink stream between BS$_i$ and its tagged user UE$_i$. The effective channel experienced by UE$_i$ is given by $h_{ii} z_{ii}^{-\alpha/2}$. Let the noise power be $\sigma_{i}^2$ and transmit power to be unity. Therefore, under total interference cancellation, the post processing SINR obtained by UE$_i$ is given by:
\begin{equation}
{SINR_{tic} = \frac{|h_{ii}|^2 z_{ii}^{-\alpha}}{\sigma_i^2} } \label{sinrtic}
\end{equation}

The CDF of rate for a UE$_i$ is obtained by the finding the limiting expression for the result \eqref{taufinal} derived in Appendix B under the limiting conditions: number of cooperating BSs $l~=1$ and interference $I_r~=0$ . On evaluating the limit,
the expression for CDF of rate under total interference cancellation ($\tau_{tic}$) when the BS intensity is $\lambda_b$ and the noise variance being $\sigma_i$ is
\begin{equation}
{\tau_{tic}\left(\lambda_b, \sigma_i, t \right)= 2 \pi \lambda_b \int_{z>0} e^{-z^2 \left[\lambda_b \pi + \mu \left(e^{t} -1 \right)  \sigma_i^2 \right]} z~\mathrm{d}z} \label{tautic}
\end{equation}

\begin{figure}[!t]
\centering
\includegraphics[width =5in]{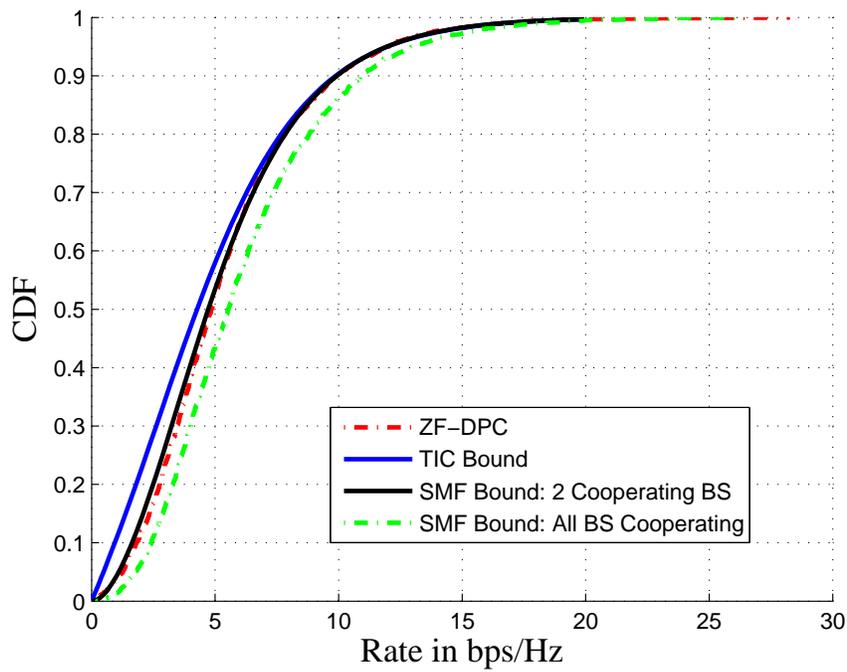}
\caption{Statistics of bounds on capacity for ZF-DPC; operating $SNR~= 10~dB$}
\label{bounds}
\end{figure}

Fig \ref{bounds} shows the relation between the CDF of achievable rate through ZF-DPC and the CDF of the total interference cancellation bound. At low rates, the difference is about $3.5~dB$ whereas at high rates the bound is tight. Therefore, for cell-edge UEs, the performance obtained through ZF-DPC is better than that achieved through total interference cancellation.

\subsubsection{Spatial Match Filter Bound}
The spatial match filter bound is obtained by match filter precoding of multiple streams of the interference terms from OCI. The post processing SINR at UE$_j$ with ${k}$ interfering BSs is given by
\begin{equation}
{SINR_{smf} = \frac{1}{\sigma_j^2} \sum_{\substack{ i \in \{1,2,\hdots,k\}}} | h_{ij}~z_{ij}^{\alpha /2}|^2} \label{sinrsmf}
\end{equation}

For simplicity, using the PPP framework, we derive the expression for CDF of rate under spatial match filer processing for a cloud of 2 cooperating BSs. The expression for coverage of rate is obtained by letting the interference term to zero in the main result from Appendix B. The expression for coverage of rate is as follows and the value is computed using numerical integration of the relation given below.
\begin{align}
{\tau_{smf}(\lambda_b,\sigma_i,t)} &= (2\pi\lambda_b)^2 \int_{z_{1i}>0} z_{1i} \int_{z_{2i}>0}  \frac{z_{1j}^\alpha - z_{2j}^ \alpha} {z_{2i} e^{- \pi \lambda_b z_{2i}^2}} \left[ A(z_{1i},z_{2i})  - A(z_{2i},z_{1i}) \right] {~\mathrm{d}z_{1i} ~\mathrm{d}z_{2i}} \label{taufinal0} \\
\intertext{where,}
A(x,y) &= x^\alpha e^{-y^\alpha \mu ( e^t - 1 ) \sigma_i^2} \nonumber
\end{align}

Fig \ref{bounds} also shows the comparison between the CDF of rate for a network using ZF-DPC and the spatial match filter upper bound. If the CSI from only two channels is used for match filter combining instead of  the information from all channels, the bound obtained is just equal to the achieving rate curve of ZF-DPC with all channels used for precoding. On the other hand, if all channels were combined in a spatial matched filter fashion, then the bound obtained is within $0.7~dB$ of the rate CDF of ZF-DPC. This close proximity between rate region due to low complexity ZF-DPC precoding and the match filter processing shows the ZF-DPC is a good approximation of the best possible scheme capturing about $85~\%$ of the theoretically achievable rate.

\section{Uplink Channel Capacity}
By the uplink-downlink duality discussed in Section II, we know that capacity for a ZF-DPC based cloud radio network operating in downlink will be same as that in uplink. However, in the downlink, as the BSs in the cloud can cooperate and jointly decode the streams, complex DP precoding is not necessary. Successive cancellation of interference using previously decoded symbols is sufficient to achieve rates similar to that achievable by DPC in downlink. This makes cloud radio operation comparatively more realizable in uplink than in downlink.

At the transmitter, a cloud radio operating in uplink uses a QR-based decoder as discussed in Appendix A with the only variation that  the matrix used for decomposition is ${H^T}$ because the channel matrices for uplink and downlink are transposes of each other. At the receiver, the cloud of BSs cooperate to sequentially decode the received symbols.

\begin{figure}[!t]
\centering
\includegraphics[width =5in]{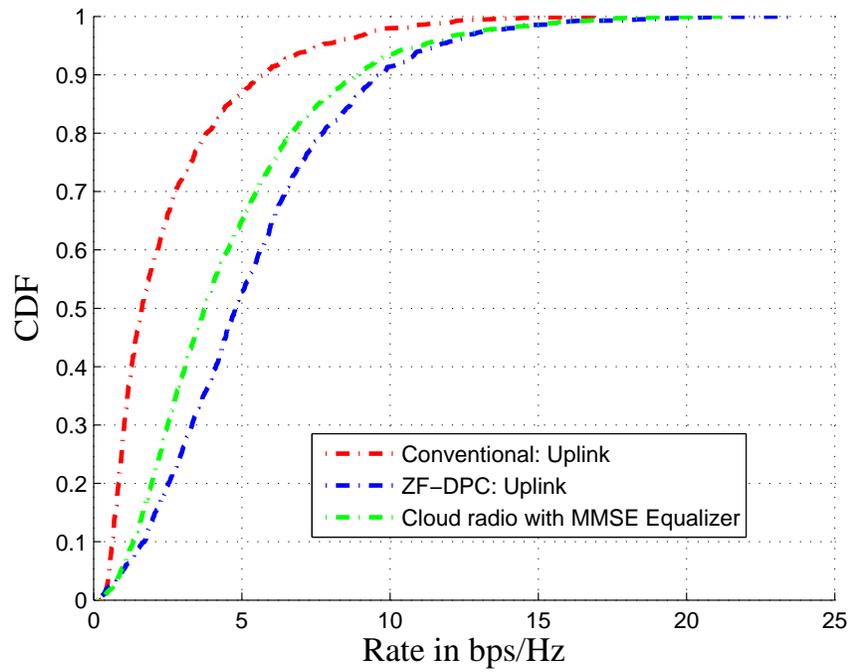}
\caption{Uplink and Downlink rate CDF for a cloud radio; operating $SNR~= 10~dB$}
\label{uplink}
\end{figure}

We obtain the achievable rate for uplink by simulating a cloud radio network in uplink and applying the aforementioned decoding to obtain the statistics of capacity. Fig \ref{uplink} compares that rate statistics of a cloud radio operating in uplink with that of the one in downlink. As expected from the uplink-downlink duality, the statistics of achievable rates are quite similar for uplink and downlink. The BSs and UEs enjoy a high increase in rate for both uplink and downlink, which makes ZF-DPC a good choice for cloud radio operation.

As the BSs can cooperate to jointly decode the received symbols, other joint decoding schemes can also be applied. We have used a low-complexity MMSE equalizer to jointly decode the received symbols. The MMSE equalizer attempts to minimize the mean squared error (MSE) and the capacity for the i$^th$ stream sent from UE$_i$ is obtained as \cite{AlDhahir:Oct00}:
\begin{align}
\textbf{\texttt{MSE}} &= \sigma^2*\left( H^\dagger H + \sigma^2*I \right)^{-1} \nonumber \\
\Rightarrow {C_{i,mmse}} &= {\log~| \texttt{diag}_i~ \left( \textbf{\texttt{MSE}} \right)|^{-1}} \label{cmmse}
\end{align}
where $\texttt{diag}_i$ denotes the $i$th diagonal entry.

Fig \ref{uplink} shows the rate statistics when an MMSE equalizer is used. Using an MMSE equalizer in a cloud radio in uplink gives upto $128~\%$ increase in the mean achievable rate as obtained by a conventional cellular network, causing an increase from $1.63~bps/Hz$ to $3.73~bps/Hz$. The cell-edge UEs in a cloud radio network with MMSE equalizer in uplink enjoy a $71~\%$ increase in capacity as compared to the conventional cellular network. From the plot we also notice that an MMSE based cloud radio captures up to $75~\%$ of the capacity achievable through a ZF-DPC based cloud radio. Therefore, the low complexity and high gain in rate, makes MMSE a good choice for uplink processing in a cloud radio. In the uplink, power control can be used to further improve the achievable rate. However, as discussed in the downlink section, the minor gain through power control as with water-filling in downlink, at the cost of increase in complexity of the system will cause an overall complexity increase in the system at the cost of very little gain improvement.

\section{Capacity with Partial CSI}
As the size of the cloud (${k}$) increases, the amount of CSI required for precoding increases as ${O(k^2)}$. For instance, in a cloud with ${k~=2}$, the number of individual channel states required is 4, whereas this number increases to 900 for a cloud with 30 BSs. Obtaining such high CSI in a cellular network is practically not feasible. Therefore, there arises a need to explore the effect of partial CSI on the achievable rate using ZF-DPC. In practice, channel states of at least 6 best BSs are estimated in conventional cellular networks.

In the next subsection, we discuss what fraction of the ZF-DPC rate can be achieved with limited channel feedback. Then we study the theoretical bounds on the capacity for processing with partial CSI. Then we present a practical scenario of partial CSI by limiting the cloud size using clustering of near-by BSs.

\subsection{ZF-DPC with partial CSI}
\begin{figure}[!t]
\centering
\includegraphics[width =5in]{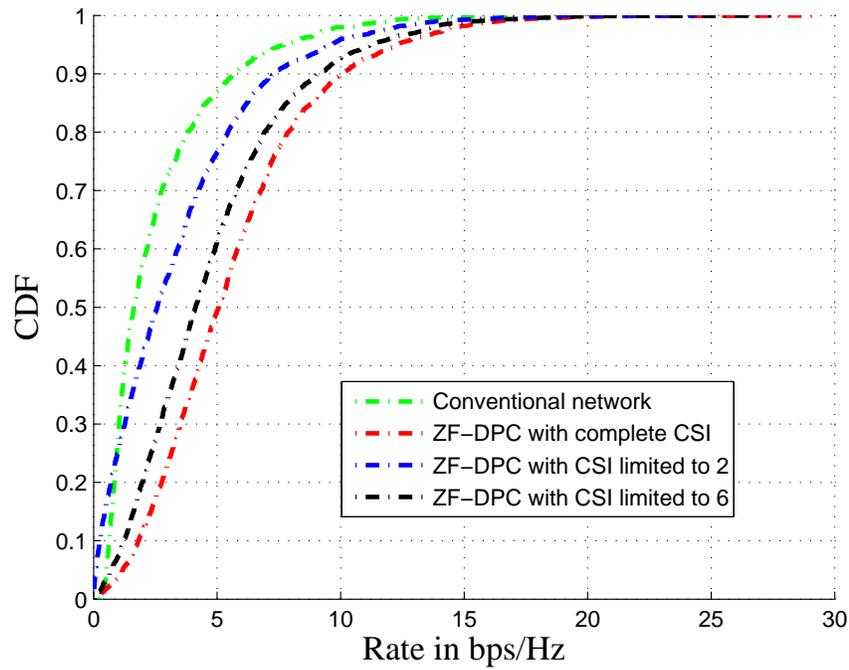}
\caption{Achievable rate statistics for ZF-DPC with complete and partial CSI; operating $SNR~= 10~dB$}
\label{partial}
\end{figure}

The channel matrix with partial CSI ${H_p}$ can be factorized as a product of ${R_p}$ and ${Q_p}$ using QR-decomposition. Using ${Q_p}$ as the precoding matrix and the channel matrix being ${H~= RQ}$ we obtain,
\begin{equation}
{Y = R Q Q_p X + N} \label{parsys}
\end{equation}
But ${QQ_p \neq I}$, leaving behind a non-upper triangular matrix for DP precoding. To further reduce the magnitude of off diagonal elements, we use $Z~= HQ_p- R_p$ for decoding instead of using $HQ_p$. Although the complexity associated with the channel feedback is reduced by limiting CSI, the complexity of the DP precoder increases. Secondly, only a fraction of the total achievable rate obtained for ZF-DPC with complete CSI is captured with partial CSI. Fig \ref{partial} shows the CDF of achievable rate for ZF-DPC with limited CSI.

When the CSI is limited to 2 channels, a cell-edge UE in a cloud radio with ZF-DPC suffers from a rate penalty as compared with a conventional network, even though the mean rate for that network experiences is increased by $48~\%$. However, current networks can support CSI of alt least up to 6 channels \cite{cdma}. In such a case, the cell-edge UE experiences a $30~\%$ increase in the achievable rate as compared to a conventional network whereas the mean achievable rate of the network increased by $140~\%$. Further rate improvements are feasible by increasing the operating SNR. Hence, ZF-DPC is an ideal choice for could radio network operation in the downlink due to a high gain in capacity even with limited feedback as per conventional standards. Furthermore, limiting the CSI to 6 channels captures about $81~\%$ of the rate that could be achieved with complete CSI which is further motivation to use ZF-DPC in the downlink.

\subsection{Theoretical bounds for partial CSI}
With partial CSI, total interference cancellation cannot be achieved. On the other hand, the CSI of known streams can be combined using spatial match filter processing. For a network with CSI limited to ${l}$ channels, the post processing SINR for UE$_j$ is given by:
\begin{equation}
{SINR_{p} = \frac{\sum_{i~= 1}^{l} | h_{ij}~z_{ij}^{\alpha /2}|^2}{\sigma_j^2 + I_r}} \label{sinrp}
\end{equation}
where ${I_r~= \sum_{{i  \in  \Phi_b \setminus \{b_1,b_2,\hdots,b_l\}}} | h_{ij}~z_{ij}^{\alpha /2}|^2}$. The evaluation of rate CDF is discussed in Appendix B. We have presented the derivation for the probability of coverage of rate for ${l~=2}$, using which the CDF of rate can be evaluated through numerical integration. The expression for CDF of rate for $l~=2$ is given as:
\begin{align}
{\tau_{smf}(\lambda_b,\sigma_i,t)}
&=(2\pi\lambda_b)^2 \int_{z_{1i}>0} z_{1i} \int_{z_{2i}>0}  \frac{1}{z_{1j}^\alpha - z_{2j}^ \alpha} ~z_{2i} e^{- \pi \lambda_b z_{2i}^2} \nonumber \\
&~~~~\left[ A(z_{1i},z_{2i}) \mathcal{L}_{I_r}(z_{2i}^\alpha \mu (e^t -1 )) - A(z_{2i},z_{1i}) \mathcal{L}_{I_r}(z_{1i}^\alpha \mu (e^t -1 ))\right] {~\mathrm{d}z_{1i} ~\mathrm{d}z_{2i}}\label{taufinal}
\intertext{where,}
A(x,y) &= x^\alpha e^{-y^\alpha \mu ( e^t - 1 ) \sigma_i^2} \nonumber \\
\mathcal{L}_{I_r}\left({y^\alpha \mu \left( e^t - 1 \right)}\right)&= exp\left[-2 \pi \lambda_b \int_{z}^{\infty} \frac{(e^t -1) v}{(e^t -1) + v^\alpha z^{\alpha} } ~\mathrm{d}v \right] \label{lap} \nonumber
\end{align}

\begin{figure}[!t]
\centering
\includegraphics[width =5in]{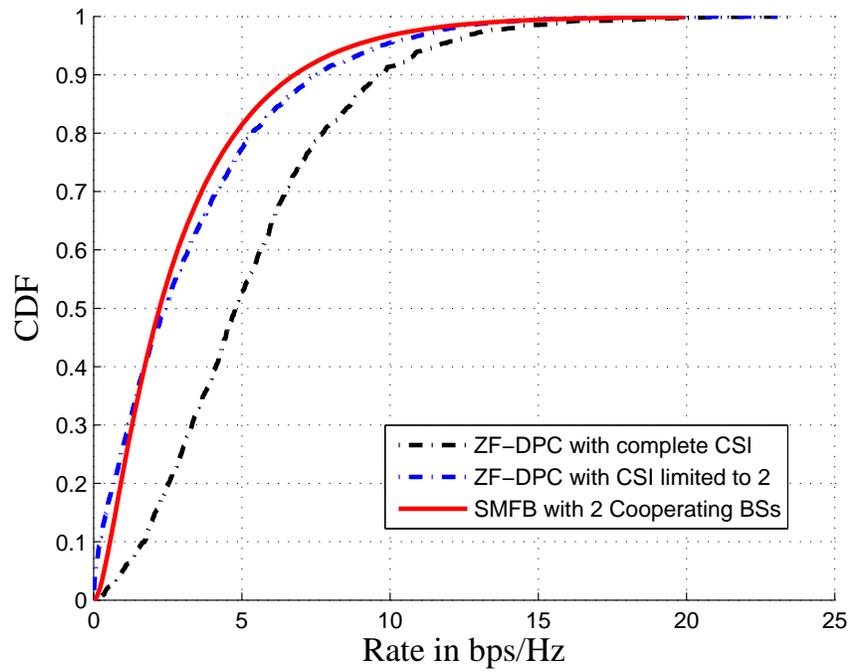}
\caption{Bounds on achievable rate with partial CSI; operating $SNR~= 10~dB$}
\label{pbound}
\end{figure}

Fig \ref{pbound} shows the rate CDF of rate for ZF-DPC based cloud radio with partial CSI and the corresponding spatial match filter bound. We observe that the spatial match filter bound for two cooperating BS is very close to that obtained by ZF-DPC with CSI limited to 2 channels. This can be further extended for any general number of cooperating BSs.

\subsection{Capacity with Geographic Clustering}
The BSs in a cloud radio network are connected to each other through optical fibers. In theory, the size of the cloud in a cloud radio network is limitless. However in real life, the size of the cloud is limited by the propagation delay in optical fiber communication. These limitations on size make geographic clustering a close approximation to a real life cloud radio where the cluster size is limited by the distances between the BS sites. Upon clustering in a cloud radio network, the effects of inter-cluster interference creep in and the achievable rate as obtained by ZF-DPC with complete CSI takes a hit because inter-cluster interference cannot be suppressed. The knowledge of the CSI for the links between BSs and UEs within  the cluster are used as input for ZF-DPC whereas the interference is due to the BSs in other clusters.

\begin{figure}[!t]
\centering
\includegraphics[width =5in]{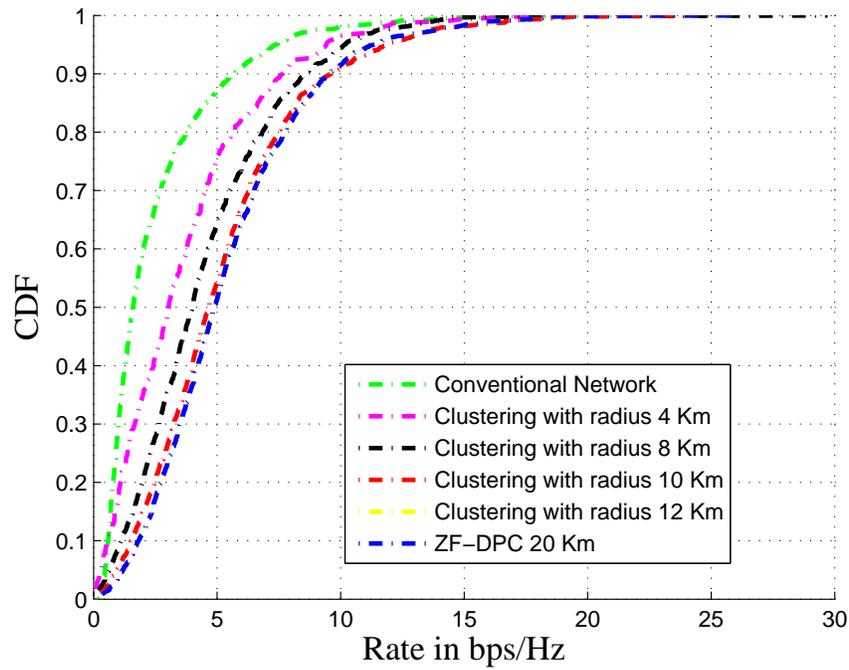}
\caption{Variations in rate statistics for a cloud radio network with limited cloud size as a function of cluster radius; operating $SNR~= 10~dB$}
\label{cluster}
\end{figure}

We have simulated a cloud radio network with clustering and studied the effect of varying clustering size on the achievable rate. Fig \ref{cluster} shows the CDF of rate achieved by a cloud radio network with limited cluster radius. An increase in cluster radius shows a relative increase in the achievable rate. For a cloud radio network with cluster radius $4~Km$, the mean achievable rate with geographic clustering is about $92~\%$ higher than that of the conventional network whereas the increase is about $193~\%$ for a cloud with cluster radius $10~Km$. A cell-edge UE in a cloud radio network with cluster radius $8~Km$ observes $41~\%$ gain in rate whereas a similar UE in a network with cluster radius $10~Km$ observes a $105~\%$ gain in rate. Depending upon the infrastructure at hand, a cloud radio network can be designed to attain any fraction of the rate achievable by ZF-DPC based cloud radio network with infinitely large cluster size. A network with cluster radius $12~Km$ achieves the same rate as that achievable through a ZF-DPC based cloud radio network with infinite cloud size.

\begin{figure}[!t]
\centering
\includegraphics[width =5in]{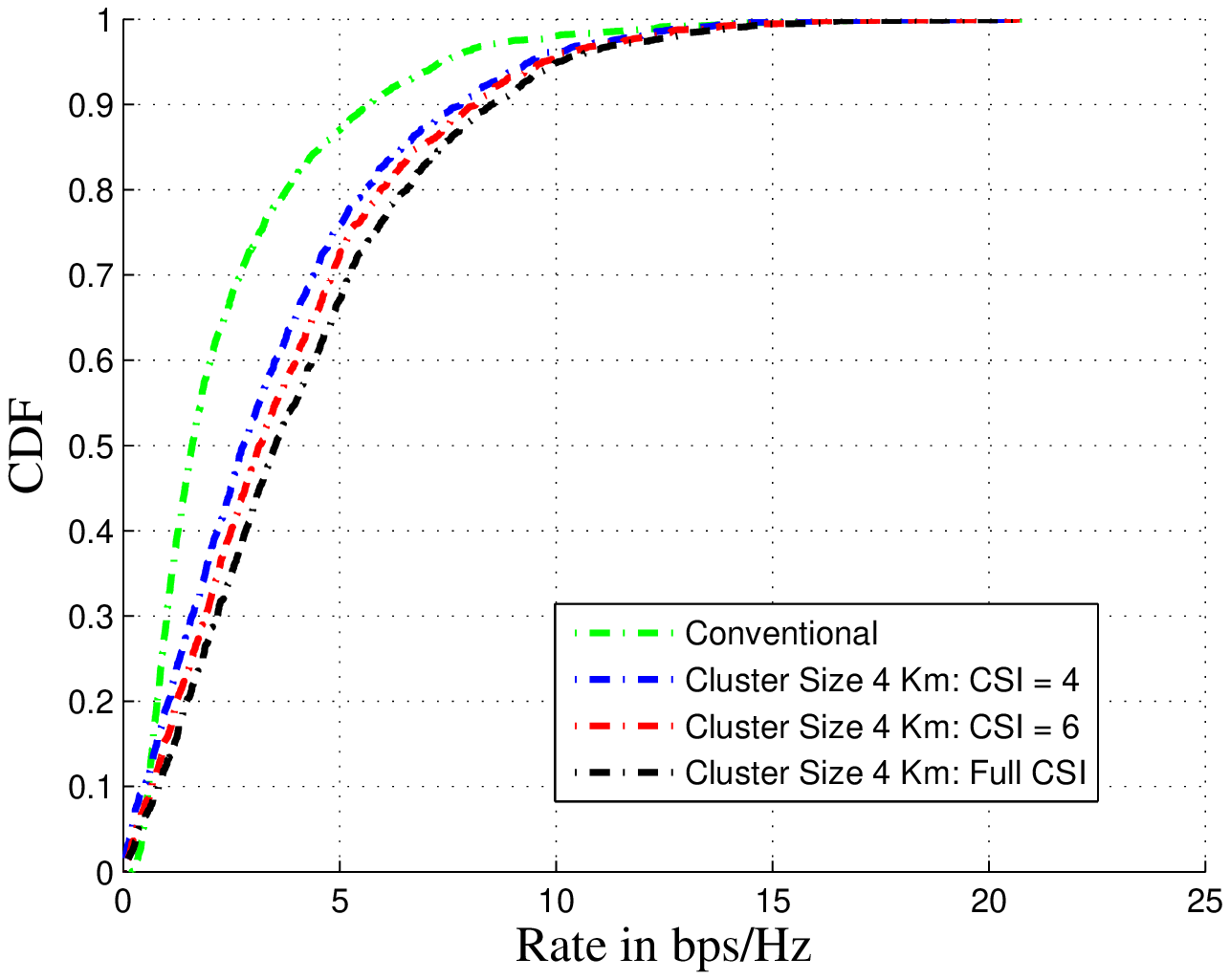}
\caption{Effect of partial CSI the rate CDF in a network with cluster radius 4 Km; operating $SNR~= 10~dB$}
\label{partial_4}
\end{figure}

\begin{figure}[!t]
\centering
\includegraphics[width =5in]{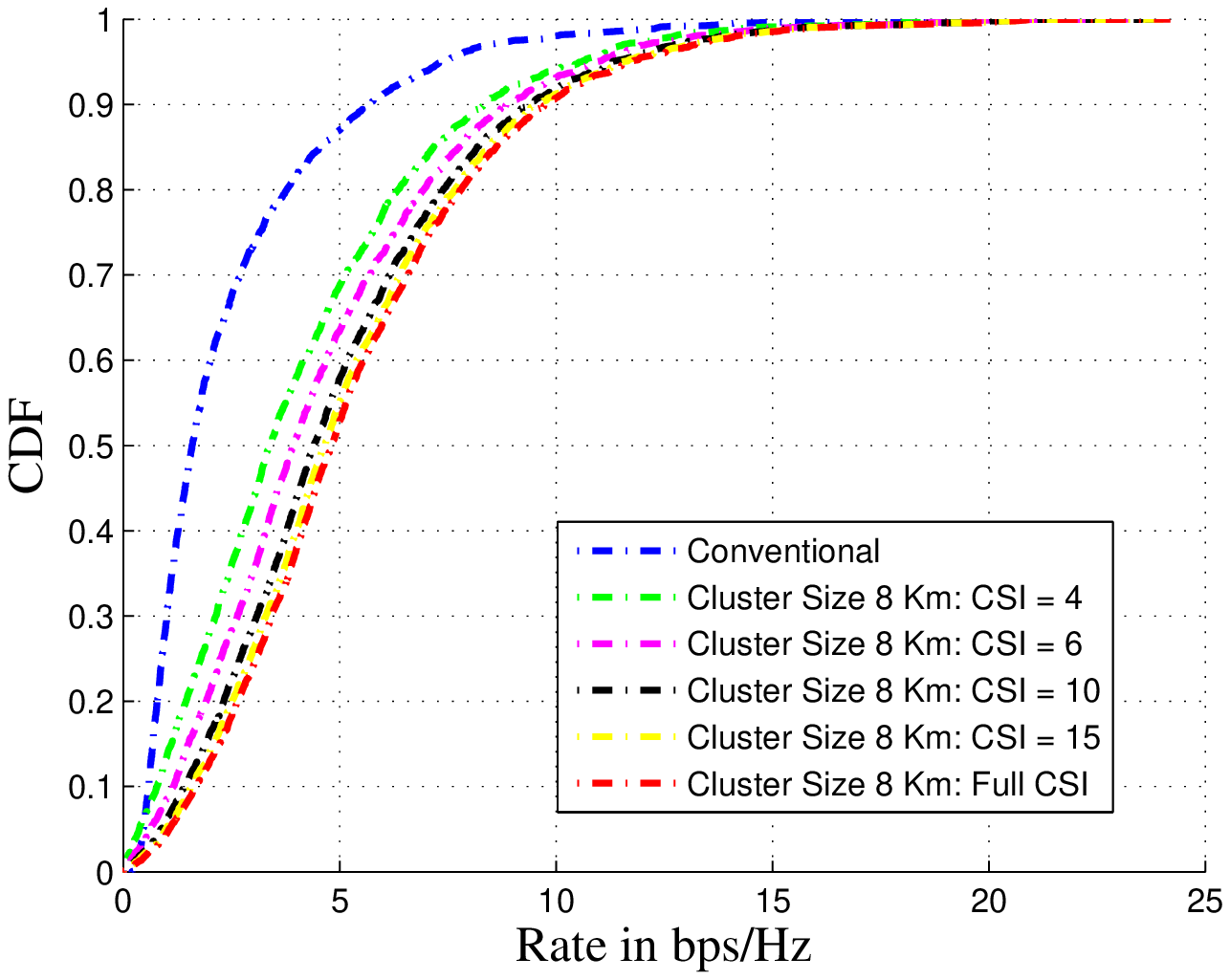}
\caption{Effect of partial CSI the rate CDF in a network with cluster radius 8 Km; operating $SNR~= 10~dB$}
\label{partial_8}
\end{figure}

We have also studied the effect of limited CSI on the achievable rate for a network with finite cluster radius. In a PPP space, the number of points in any region is dependent solely on the area of the region and is independent of the shape and location of the region in the space \cite{ganti-spcom}. In a square region of edge-length $20~Km$ the average number of points for a PPP with intensity $0.3$ is $120$. The average number of points in clusters with radii $4~Km$ and $8~Km$ are $15$ and $60$ respectively and represent the average number of BSs in each of the clusters. Fig \ref{partial_4} shows the CDF of rate achievable by ZF-DPC with partial CSI in a cloud radio network with cluster size $4~Km$. Limiting the feedback to 4 channels captures about $78~\%$ of the average rate achievable through processing with complete CSI, where limiting the CSI to 6 captures about $89~\%$ of the rate obtainable through complete CSI. We observe that limiting the CSI to just 6 states in a cloud radio network with cluster size $4~Km$ gives a two-fold increase in mean achievable rate. However in a network with larger cluster radius, more channel states are required to capture the same fraction of the total achievable rate as shown in Fig \ref{partial_8}. To capture $90 \%$ of the mean achievable rate through processing complete CSI in a network with cloud radius $8~Km$, the knowledge of 10 channels is required whereas the knowledge of 15 channels states achieves $99 \%$ of the mean rate for complete CSI. Although more CSI is required to achieve the same fraction of the total rate for a network with larger cluster size, the associated gain in rate is also quite high. The cell-edge rate with CSI limited to 10 states is $89~\%$ higher than the conventional network and the mean achievable rate is about $175~\%$ higher as opposed to the $100~\%$ increase through a cluster of size $4~Km$.

\begin{figure}[!t]
\centering
\includegraphics[width =5in]{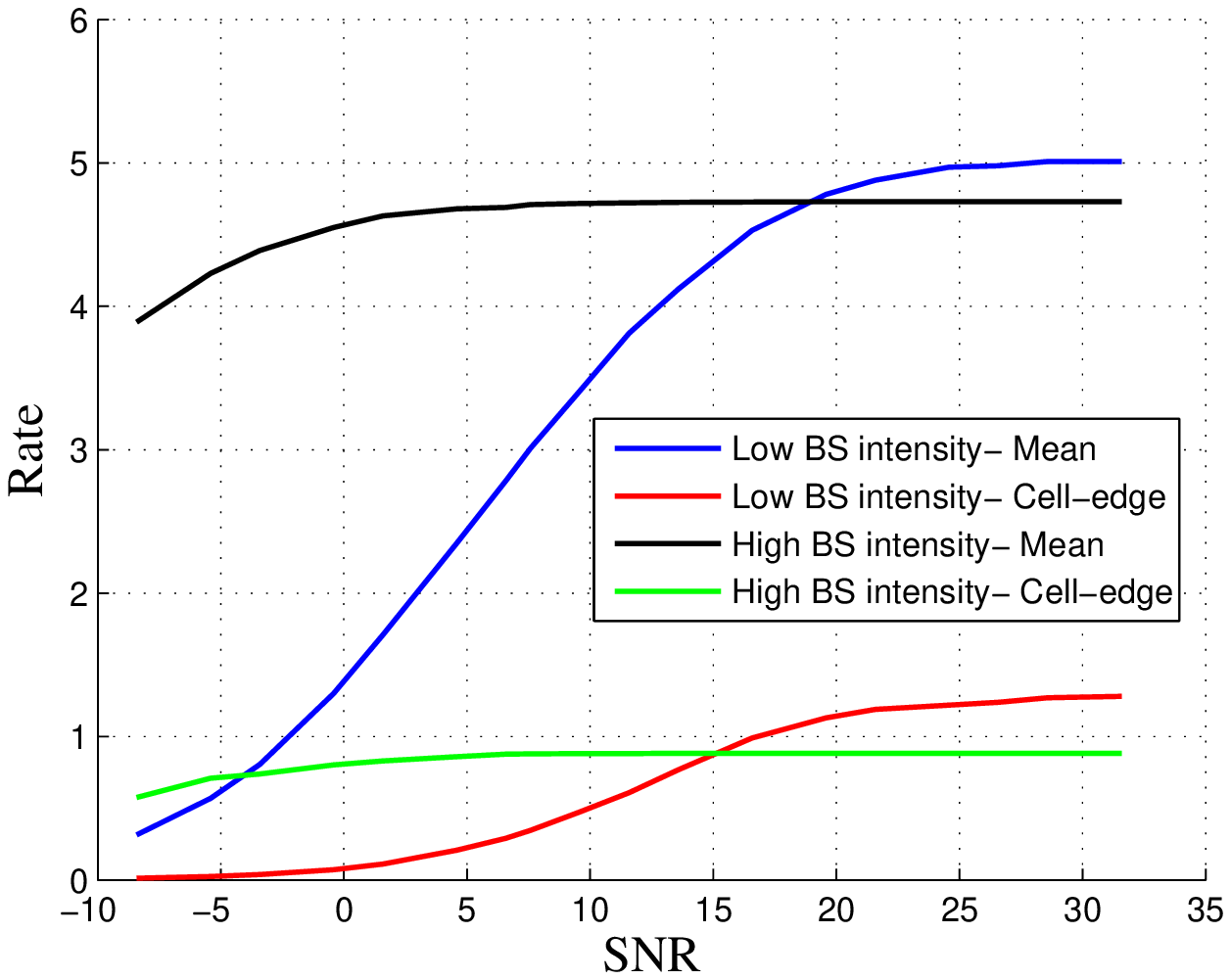}
\caption{Spectral efficiency of a ZF-DPC based cloud radio network with cluster radius $8~Km$ and CSI = 6}
\label{8_final}
\end{figure}

\begin{figure}[!t]
\centering
\includegraphics[width =5in]{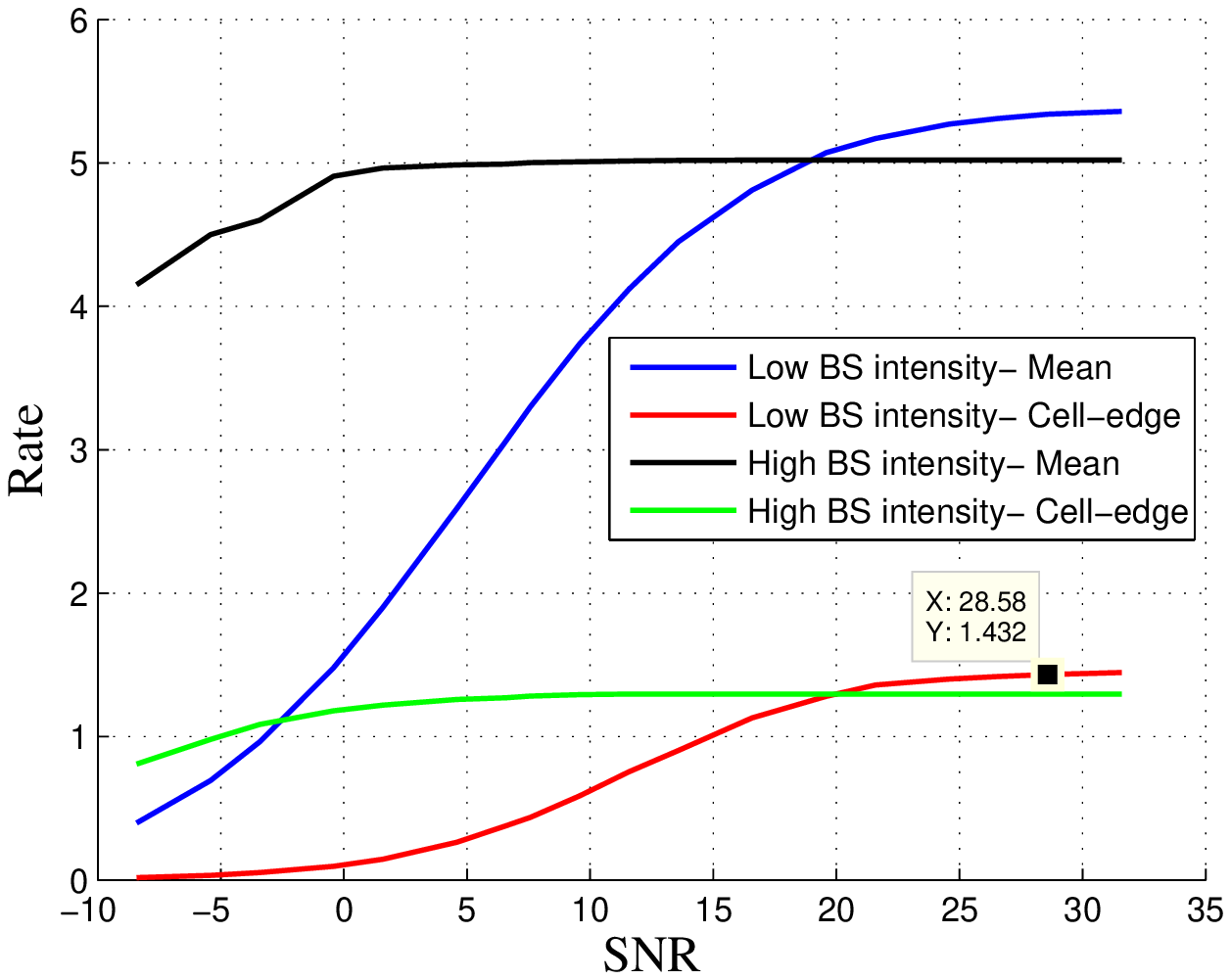}
\caption{Spectral efficiency of a ZF-DPC based cloud radio network with cluster radius $12~Km$ and CSI = 10}
\label{12_final}
\end{figure}

Lastly, the effect of variations in the operating SNR on the achievable rate has been studied to understand the noise limited/interference limited behavior of a cloud radio network. As discussed earlier, a cloud radio network with geographical clustering and limited feedback is a close approximation to the real-life cloud radio networks. In \cite{heath}, the authors have shown that a cloud radio network shows improvements in achievable rate with increase in operating SNR, in low and medium SNR regimes. However, in high SNR regimes, the achievable rate settles to a saturation point. In other words, a cloud radio network behaves as a noise limited network in low and medium SNR regimes and as an interference limited network in high SNR regimes. We have observed a similar performance for a ZF-DPC based cloud radio network with geographic clustering and limited feedback as shown in the spectral efficiency plots: Fig \ref{8_final}, Fig \ref{12_final}.

Fig \ref{8_final} shows the spectral efficiency plot of a ZF-DPC based cloud radio network with cluster radius 8 Km and CSI limited to 6 states. At low BS intensity, the mean achievable rate shows a steady increase with an increase in operating SNR upto $25~dB$, after which there is no change in the achievable rate with further increase in operating SNR. The mean achievable rate settles to a saturation value of $5.01~bps/Hz$. Similarly, the achievable rate for a cell-edge user saturates to $1.28~bps/Hz$. Therefore, the maximum possible gain through a ZF-DPC based cloud radio with cluster size $8~Km$ and CSI limited to 6 as opposed to a conventional cellular network is $211.1~\%$ and $150.9~\%$ in mean achievable rate and cell-edge achievable rate respectively.

The performance of a network with higher BS intensity shows similar characteristics except for the fact that the saturating rate is lower than that for a network with low BS intensity. The reason for loss at high intensity is as follows. For a given cluster radius, increase of BS density increases the number of BSs within the cluster. Therefore, to maintain the performance comparable to that of a network with low BS intensity, the amount of feedback needs to be increased proportionately. The maximum achievable gain for a ZF-DPC based network with high BS intensity with $8~Km$ cluster radius and CSI limited to 6, as compared with a conventional cellular network is $193~\%$, $75~\%$ in mean and cell-edge respectively.

Fig \ref{12_final} shows that a cloud radio network with a larger cluster size and increased CSI also shows similar performance. As expected the gain in mean  achievable rate and cell-edge achievable is higher than a network with smaller cluster size and reduced CSI. The gain of a ZF-DPC based cloud radio network with $12~Km$ cluster radius and CSI limited to 10 states, over a conventional network is about $230~\%$ in mean and $180~\%$ in cell-edge.\\

\section{Tomlinson-Harashima Precoder}
\begin{figure}[!t]
\centering
\includegraphics[width =5in]{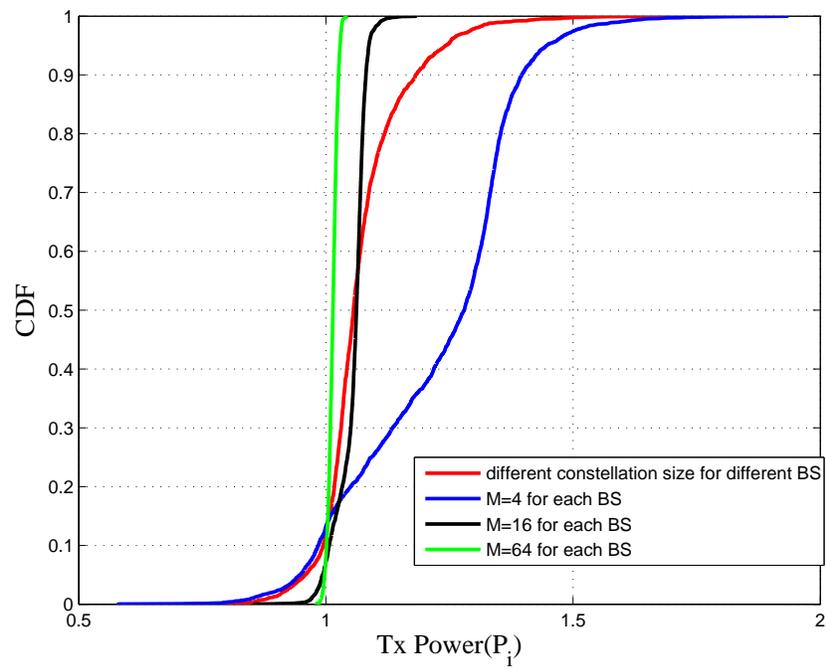}
\caption{Transmitted signal Power CDF Plot using THP; operating $SNR~= 10~dB$}
\label{Tx_pow}
\end{figure}

DPC at the transmitter side is quite similar to decision feedback equalzier (DFE). It is this combination of DFE and symmetric-modulo operation at the transmitter side that gives the THP. The modulo arithmetic is employed which bounds the symbol range, thereby reducing some or most of the increased transmit power. We have simulated a cloud radio network using a multi-user THP using the PPP framework. We use a rectangular area of 10Km with ${\lambda_b=0.3}$. The individual SNR at the transmitter for each BS in the cloud (${1/{{\sigma}_i}^2}$) is taken to be 10 dB.  The complex ${M_i}$-ary QAM is used for each BS ${i}$ and the following criteria defines the constellation size, ${M_i}$ for each BS
\[ M_i = \left\{
  \begin{array}{l l}
    64 & \quad \text{if $C_{ZFDPC}>7$}\\
    16 & \quad \text{if $4<C_{ZFDPC}\le{7}$}\\
    4 & \quad \text{if $C_{ZFDPC}\le{4}$}
  \end{array} \right.\]
where $C_{ZFDPC}$ represents the ZF-DPC capacity of each stream. In an actual system, each BS encodes the data using a capacity achieving modulation and coding scheme before the data is precoded using the THP. Here, we restrcited the constellation size to three values that are used in typical systems. Figure \ref{Tx_pow}, shows the cdf of transmit power for a cloud radio employing a fixed value of $M$ as well as for the case of adaptive modulation based on the aforementioned criterion. Note that each BS uses unit power the case of ZF-DPC. The transmit power penalty is high only when all the BS use $M=4$. However, we see that the increase in the power is quite small when the BS uses varying values of $M$. Since, like ZF-DPC, THP achieves total interference elimination with a small transmit power penalty, we can expect this method to offer a rate close to that of ZF-DPC.

\section{Conclusion}
In this paper, we have presented techniques that achieve rates close to the capacity of a cloud radio. We introduced a cloud radio network in the stochastic geometric framework using PPPs which made the rate analysis much simpler when compared to the grid based analysis techniques. Then we presented a zero forcing based DP precoder (ZF-DPC) which showed substantial improvements in the mean achievable rates as compared with conventional cellular network, at typical SNRs. We also showed that the capacity achieved by a cloud radio using ZF-DPC lies between and is quite close to the two theoretical bounds i.e the lower bound being the total interference cancellation bound and the upper bound being the spatial match filer bound. The proximity with the spatial match filter bound makes ZF-DPC a good choice for cloud radio operation due to its reduced complexity. Using the uplink-downlink duality between MAC and BC we studied the statistics of the achievable rates by a cloud radio in uplink and downlink. In the uplink, we also studied the use of an MMSE receiver instead of a QR based decoding. We found that MMSE captures upto $75~\%$ of the rate obtained by ZF-DPC.

In the latter half of the paper, we studied the impact of limited CSI on the rate using ZF-DPC based processing. We show that with CSI from 6 channels, we can capture up to $85~\%$ of the rate as obtainable using ZF-DPC with complete CSI. We verified the optimality of the performance of ZF-DPC with partial CSI by comparison with theoretical bounds. These results suggest that a substantial portion of the theoretical gain offered by cloud radio can be realized in hot spot areas where operators deploy BSs in isolated clusters of small size that permits full cooperation between the BSs.\\

We also presented the physical scenario of geographic clustering to limit the cloud size where working with partial CSI is necessary. We observed that a ZF-DPC based cloud radio network with geographic clustering and limited feedback behaves like a noise limited network in low and medium SNR regimes whereas the performance in high SNR regimes is interference limited. This result is consistent with the observation of \cite{heath}. Next, we presented a practically realizable processing scheme using zero forcing equalizer and used a THP to realize DP precoding. We have seen that THP comes very close to achieving the capacity of a DPC precoder but an additional penalty in transmit power is incurred with the use of THP.\\

The results of this paper show that ZF-DPC or THP is a promising option for implementation in a cloud radio network. However, before the actual implementation, there are other related problems which need to be explored. Specifically, it is important to characterize the number of channels that be fed-back in a cloud radio network. Further work should study pilot design aspects in detail. Additionally, techniques to handle the adverse effects of inter-cluster interference need to be developed further.

\bibliographystyle{IEEEtran} 
\bibliography{IEEEabrv,IEEEexample}

\section*{Appendix A \\ Proof of ZF-DPC}
Consider a cloud radio network with ${Y}={HWX+N}$, where ${W}$ is the precoding matrix chosen such that interference cancellation is achieved. The channel matrix ${H}$ is decomposed by QR-factorization as ${H}={RQ}$, where ${Q}$ is a unitary matrix and ${R}$ is an upper triangular matrix. When the precoding matrix is chosen as ${W} = {Q^\dagger}$ we have,
\begin{align}
{Y} &= {H Q^\dagger X + N} \label{presys} \\
\Rightarrow {Y} &= {R Q Q^\dagger X + N} \\
\Rightarrow \begin{bmatrix}
  y_1\\
  y_2\\
  \vdots\\
  y_k
 \end{bmatrix} &= \left[ \begin{array}{cccc}
  r_{11}&0&\hdots&0\\
  r_{12}&r_{22}&\hdots&0\\
  \vdots&\vdots&\vdots&\vdots\\
  r_{1k}&r_{2k}&\hdots&r_{kk}
 \end{array} \right] \times \left[ \begin{array}{c}   x_1\\
  x_2\\
  \vdots\\
  x_k
 \end{array} \right] + \left[ \begin{array}{c}   n_1\\
  n_2\\
  \vdots\\
  n_k
 \end{array} \right] \label{presys2}
 \end{align}
By DP precoding theorem, the interference terms can be completely eliminated at the transmitter.  Therefore, the post processing SINR for the UE$_i$ is given by
\begin{equation}
{SINR_i} = {\frac{|r_{ii}|^2|x_i|^2}{\sigma_i^2}} \label{sinrdpc}
\end{equation}
and the rate for UE$_i$ is given by
\begin{equation}
{C_i} = {\log\left( 1+ \frac{|r_{ii}|^2|x_i|^2}{\sigma_i^2} \right)}~bps/Hz \label{cidpc}
\end{equation}

\section*{Appendix B \\ Rate Coverage for Spatial Match Filter Processing}
We use the system model as discussed in Section II and assume unit transmit power by all BSs. For a network with $l$  cooperating BS and match filter processing, the post processing SINR is given by:
\begin{equation}
{SINR_{p} = \frac{\sum_{i~= 1}^{l} | h_{ij}~z_{ij}^{\alpha /2}|^2}{\sigma_j^2 + I_r}} \label{snrderi}
\end{equation}
where ${I_r~= \sum_{{i  \in  \Phi_b \setminus \{b_1,b_2,\hdots,b_l\}}} | h_{ij}~z_{ij}^{\alpha /2}|^2}$. For $l~= 2$, the  coverage probability for a target rate ${t}$ for UE$_i$ centred at origin is given by:
\begin{align}
{\tau_{smf}(\lambda_b,\sigma_i,t)} &= \mathbb{P}\left[ {\log}\left(1 + SINR \right) > t | z_{1i}, z_{2i} \right] \label{taudef}\\
&= \mathbb{P}[
| h_{1i}|^2~\mathrm{d}_{1i}^{\alpha} +
| h_{2i}|^2~\mathrm{d}_{2i}^{\alpha} > \left( e^t - 1 \right) (I_r + \sigma_i^2)| z_{1i}, z_{2i}] \label{tauprob}
\end{align}
where ${h_{1i}, h_{2i}} \sim \mathcal{CN}(0,\frac{1}{\mu})$. In the PPP framework, the null probability of BS distribution in a region of radius $z_{1i}$ centred at origin is
\begin{align}
\mathbb P[z > z_{1i}] &= \mathbb P [No~BS~closer~than~distance~z_{1i}] \nonumber\\
&= e^{-\lambda_b \pi z_{1i}^2} \label{null}\\
\intertext{Therefore,the CDF is}
\mathbb{P}[z < z_{1i}] &= F_1(z_{1i}) \nonumber \\
&= 1- e^{-\lambda_b \pi z_{1i}^2}
\end{align}
Hence, the distributions of distance ${z_{1i}}$ follow as
\begin{align}
{z_{1i}} \sim f_1(z_{1i})&= 2 \pi \lambda_b z_{1i} e^{- \pi \lambda_b z_{1i}^2} \label{z1}
\end{align}
From \cite{icc}, it is known that the probability of there being exactly $k$ BSs within two concentric circles of radii $z_{1i}$, $z_{2i}$ centred at origin $f_k(z_{2i}|z_{1i})$ is given by:

\begin{equation}
f_k(z_{2i}|z_{1i}) = \frac{2 \pi \lambda z_{2i}}{\left(k-2\right)!}~\left(\pi\lambda\left(r_2^2 - z_{1i}^2\right)\right)^{k-2}~e^{-\pi \lambda\left(z_{2i}^2 - z_{1i}^2\right)} \nonumber
\end{equation}
Therefore, for $k~= 2$ we obtain:
\begin{align}
{z_{2i}} \sim f_2(z_{2i}|z_{1i})&= 2 \pi \lambda_b z_{2i} e^{- \pi \lambda_b \left( z_{2i}^2 - z_{2i}^2 \right)} \label{z2}
\end{align}
The linear combination of $|h_{ij}|^2$ follows hyper-exponential distribution whose tail probability is give by:
\begin{align}
G(z_{1i},z_{2i},t) &= \mathbb{P}[ | h_{1i}|^2~\mathrm{d}_{1i}^{\alpha} + | h_{2i}|^2~\mathrm{d}_{2i}^{\alpha} > \left( e^t - 1 \right) (I_r + \sigma_i^2)] \nonumber \\
&= \frac{1}{z_{1j}^\alpha - z_{2j}^ \alpha} \left[{z_{1i}^\alpha e^{-z_{2i}^\alpha \mu ( e^t - 1 ) (I_r + \sigma_i^2)} - z_{2i}^{\alpha} e^{-z_{1i}^\alpha \mu \left( e^t - 1 \right) (I_r + \sigma_i^2)}}\right] \label{hypexp}
\end{align}

The expression for the coverage of rate of a fixed target rate is given as:
\begin{equation}
{\tau_{smf}(\lambda_b,\sigma_i,t)}  = \int_{z_{1i}>0} f_1(z_{1i}) \int_{z_{2i}>0} f_2(z_{2i}|z_{1i})~G(z_{1i},z_{2i},t) {~\mathrm{d}z_{1i} ~\mathrm{d}z_{2i}} \label{tausmf2}
\end{equation}
Substituting results from \eqref{z1}, \eqref{z1} and  \eqref{hypexp}, we obtain:
\begin{align}
{\tau_{smf}(\lambda_b,\sigma_i,t)} &= \int_{z_{1i}>0} 2 \pi \lambda_b z_{1i} e^{- \pi \lambda_b z_{1i}^2} \int_{z_{2i}>0}  \frac{1}{z_{1j}^\alpha - z_{2j}^ \alpha} ~2 \pi \lambda_b z_{2i} e^{- \pi \lambda_b \left( z_{2i}^2 - z_{2i}^2 \right)} \nonumber \\
&~~~~~~~~~~~~~~~~~~~~~~~~~~~~~\left[ { z_{1i}^\alpha e^{-z_{2i}^\alpha \mu ( e^t - 1 ) (\sigma_i^2 + I_r)} - z_{2i}^{\alpha} e^{-z_{1i}^\alpha \mu \left( e^t - 1 \right) (\sigma_i^2 + I_r)}} \right]   {~\mathrm{d}z_{1i} ~\mathrm{d}z_{2i}} \label{tausmf3} \\
&=(2\pi\lambda_b)^2 \int_{z_{1i}>0} z_{1i} \int_{z_{2i}>0}  \frac{1}{z_{1j}^\alpha - z_{2j}^ \alpha} ~z_{2i} e^{- \pi \lambda_b z_{2i}^2} \nonumber \\
& ~~~~~~~\left[ A(z_{1i},z_{2i}) \mathcal{L}_{I_r}(z_{2i}^\alpha \mu (e^t -1 )) - A(z_{2i},z_{1i}) \mathcal{L}_{I_r}(z_{1i}^\alpha \mu (e^t -1 ))\right] {~\mathrm{d}z_{1i} ~\mathrm{d}z_{2i}}\label{taufinal} \\
\intertext{where,}
A(x,y) &= x^\alpha e^{-y^\alpha \mu ( e^t - 1 ) \sigma_i^2} \nonumber
\end{align}
and $\mathcal{L}_{I_r}$(s) is the Laplace transform evaluated w.r.t $I_r$. For any distance $z$, measured from origin in a PPP space, the Laplace transform w.r.t $I_r$ is obtained from \cite{andrews} as:
\begin{align}
\mathcal{L}_{I_r}\left({y^\alpha \mu \left( e^t - 1 \right)}\right)&= exp\left[-2 \pi \lambda_b \int_{z}^{\infty} \frac{(e^t -1) v}{(e^t -1) + v^\alpha z^{\alpha} } ~\mathrm{d}v \right] \label{lap}
\end{align}
The probability of coverage is obtained by performing numerical integration operations on \eqref{taufinal}. The CDF of rate for any generic cloud size, can be obtained by a simple extension of the proof by evaluating the tail probability for a sum of $k$ squared Gaussian terms instead of two in \eqref{hypexp}.


\end{document}